\documentclass[preprint2]{aastex}

\received{receipt date}
\revised{revision date}
\accepted{acceptance date}

\begin{document}

\title{Accurate parameters of the oldest known rocky-exoplanet hosting system: Kepler-10 revisited}
\author{Alexandra Fogtmann-Schulz, Brian Hinrup, Vincent Van~Eylen, J{\o}rgen Christensen-Dalsgaard, Hans Kjeldsen, V\'ictor Silva Aguirre, and Brandon Tingley}
\affil{Stellar Astrophysics Centre, Department of Physics and Astronomy, Aarhus University, Ny Munkegade 120, DK-8000 Aarhus C, Denmark.}
\email{alfosc@phys.au.dk}

\begin{abstract}
Since the discovery of Kepler-10, the system has received considerable interest because it contains a small, rocky planet which orbits the star in less than a day. The system's parameters, announced by the \textit{Kepler} team and subsequently used in further research, were based on only 5 months of data. We have reanalyzed this system using the full span of 29 months of \textit{Kepler} photometric data, and obtained improved information about its star and the planets. A detailed asteroseismic analysis of the extended time series provides a significant improvement on the stellar parameters: Not only can we state that Kepler-10 is the oldest known rocky-planet-harboring system at $10.41\pm1.36$ Gyr, but these parameters combined with improved planetary parameters from new transit fits gives us the radius of Kepler-10b to within just 125 km. A new analysis of the full planetary phase curve leads to new estimates on the planetary temperature and albedo, which remain degenerate in the \textit{Kepler} band. Our modeling suggests that the flux level during the occultation is slightly lower than at the transit wings, which would imply that the nightside of this planet has a non-negligible temperature. 
\end{abstract}
\section{Introduction}
The \textit{Kepler} mission has been continuously monitoring over $10^5$ stars since its launch in 2009. Among the more interesting targets is Kepler-10 \citep[KIC~11904151,][]{2011ApJ...729...27B}; a planet host star on the brighter end of the \textit{Kepler} sample (Kepler magnitude $Kp=10.96$), which harbors at least two planets. The innermost planet, Kepler-10b, was the first confirmed rocky planet discovered by \textit{Kepler} and orbits its star in only 0.84 days. A second transiting planet, Kepler-10c, is a super-Earth with a radius of about 2$R_\oplus$ and a 45-day period \citep{2011ApJ...729...27B}. While Kepler-10b induces detectable radial velocity variations in the host star as it orbits, Kepler-10c does not, so its mass is not known. Possible false positive scenarios for Kepler-10c have been ruled out using the BLENDER analysis \citep{torres2011} including an observed transit at 4.5~$\mu$m with the \textit{Spitzer Space Telescope} \citep{fressin2011}.

The properties of the host star were previously determined by \citet{2011ApJ...729...27B} using asteroseismology, based on five months of \textit{Kepler} data. In this paper, we revisit the asteroseismic analysis based on a dataset that is now about six times as long. We derive accurate stellar parameters; in particular, we set out to constrain the age of the star and its planets, which potentially is high \citep[$11.9\pm4.5$ Gyr,][]{2011ApJ...729...27B}. The age also enters in composition models of Kepler-10b \citep[e.g.,][]{gong2012}, which, together with CoRoT-7b \citep{leger2009}, has been the subject of many studies of planetary dynamics and migration, because of its small size and proximity to the star \citep[e.g.,][]{dong2013}. The planet is thought to have a Mercury-like composition \citep{gong2012}, with a large iron core and silicate mantle \citep{wagner2012}. The origin of the planet could be a rocky planet which has lost a thin hydrogen envelope \citep{leitzinger2011} or the remnant of a 
Jupiter-mass giant planet \citep{kurokawa2013}.

The planet's short period also results in a detectable full planetary phase curve, by measuring the increase in flux as the dayside of the tidally locked planet rotates into view, as well as an occultation, when the planet disappears behind its star. Analyzing the phase curve allows for an estimate of the planetary temperature and albedo. Together with the very recently discovered Kepler-78 \citep{sanchis2013}, this is the only terrestrial exoplanet for which this is possible. The first detection of the phase curve and occultation \citep{2011ApJ...729...27B} was later physically interpreted using a \textit{Lava-Ocean} hypothesis: a planet without atmosphere and a surface partially covered with molten rocks \citep{rouan2011}. We use the far longer time series currently available, to measure the phase curve amplitude and occultation depth at a significantly improved accuracy.

In Section~\ref{sec:asteroseismology}, we present the results of a detailed asteroseismic analysis. We use the new stellar parameters to fit the planetary transits to higher precision (Section~\ref{sec:transit}). In Section \ref{sec:phasecurve}, we analyze the amplitude of the phase curve and the depth of the occultation; these values are interpreted in terms of temperature and albedo in Section \ref{sec:albedo}.
\section{Host star asteroseismology}\label{sec:asteroseismology}
The asteroseismic analysis was based on the \textit{Kepler} raw fluxes, which were acquired using simple aperture photometry and uncorrected for any instrumental effects. Short-cadence data (which have a sampling rate of 58.8 seconds) were used up to \textit{Kepler} quarter 14, where a quarter (Q) normally contains three months of data. Due to a hardware failure in Q4, the target was only observed during its first month and subsequently not observed during Q8 and Q12, resulting in a total of 29 months of observations which were used. 

We analyzed the data from each month separately, correcting for transit-like signatures, long-term trends and bad data points (outliers). The long-term corrections were done by a median filtering algorithm which removes all trends with a time scale longer than 0.7 days. Transit-like signatures were removed by applying a median filter (with a 40-minute time scale) to all parts of the time series where the data deviate significantly from the 0.7 day median filter. Since we modeled the phase curve for the two known exoplanets in the Kepler-10b system an alternative would have been to remove the transit signatures directly after constructing the phase curve (see Section~\ref{sec:phasecurve}). We decided to use a general algorithm (the median filter described above) for removal of any transit-like signature in order not to depend on modeling the exoplanet system. The difference between the median filter and a direct removal of the phase curve for each known exoplanet is insignificant and the frequency where the p-modes are present and the individual measured p-mode frequencies do not depend on the choice of algorithm for removal of transit-like signatures and long-term trends. The noise level at high frequencies is not substantially affected  by the transits, and only the lowest dozen harmonics of the exoplanet period contains significant power. The lowest detected p-mode frequency corresponds to a period which is two orders of magnitude shorter than the period of Kepler-10b.

\begin{figure*}[ht]
 \includegraphics[width = \textwidth]{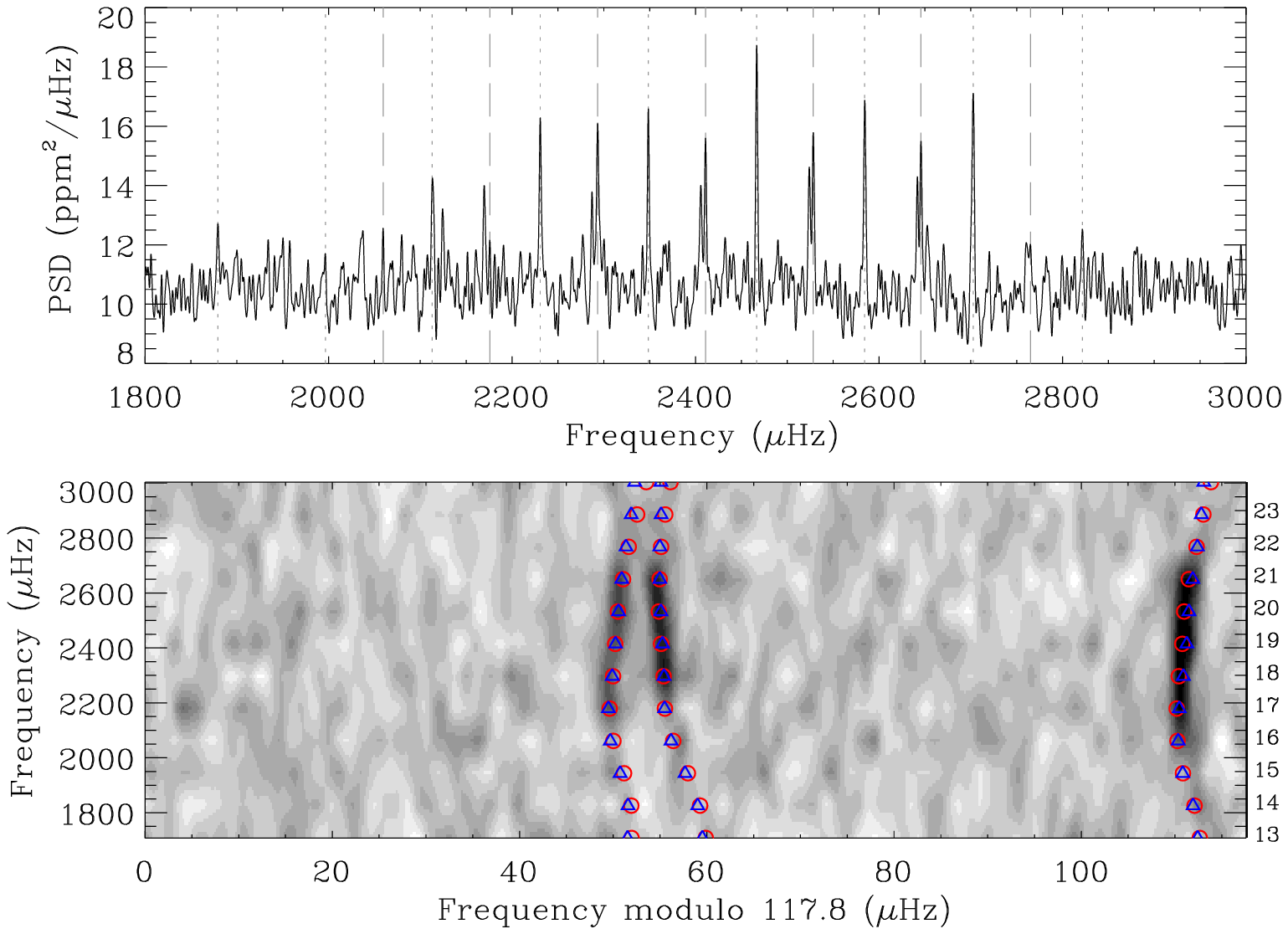}
 \caption{\textit{Top:} Power spectrum of Kepler-10. Dashed lines mark the $l=0$ modes, dotted lines mark the $l=1$ modes. The spectrum is calculated as the weighted average of 29 individual power spectra (29 month of Kepler data). The intrinsic resolution of the averaged spectrum is 0.3922 $\mu$Hz. \textit{Bottom:} \'Echelle diagram of Kepler-10, constructed by dividing the power spectrum into segments corresponding to the large frequency separation and stacking them on top of each other. Also plotted are the frequencies from the best-fit models of GARSTEC (blue open triangles) and ASTEC (red open circles), both corrected for near-surface effects.}
 \label{fig:power}
 \end{figure*}

We used the resulting differential photometry and estimated errors to obtain the power spectrum. We first calculated a weighted power spectrum for each month of data separately, with the weights proportional to the inverse of the variance. We then combined the monthly power spectra, each weighted by the square of the inverse of the median of its power spectrum. This resulting final power spectrum, which we used for the asteroseismic analysis, has a resolution of about 0.4 $\mu$Hz (Figure~\ref{fig:power}, top).

We extracted the individual oscillation peaks by calculating the power-weighted frequency position for each peak in the spectrum. From the distribution of random peaks in the spectrum, we estimated a detection threshold above which any power is likely to be real, either stellar oscillations or a coherent signal in the instrument. We identified the orders and degrees of the individual extracted p-modes from the structure of the frequencies using the large and small frequency separation. We estimated errors in frequencies using a Monte Carlo simulation where we injected signals in the time series at similar power levels to those observed, extracted them, then compared the input and output frequencies. The individual frequencies can be seen in Table~\ref{table:frequencies}. The bottom of Figure~\ref{fig:power} shows the observed oscillation frequencies in an \'echelle diagram.

Determinations of the surface abundances and effective temperature of Kepler-10 have been reported by \cite{torres2012}. The authors used high-resolution spectroscopic observations coupled to asteroseismic determinations of stellar gravity to improve the obtained accuracy in the metallicity and temperature. These are given in Table~\ref{table:parameters}.
\begin{table}[ht]
\centering
\caption{Oscillation peaks for Kepler-10.}
\label{table:frequencies}
\begin{tabular}{l l l}
\tableline\tableline
Frequency ($\mu$Hz) & $l$ & $n$ \\ 
\tableline
1879.3 $\pm$ 0.4 &    1 & 14 \\
1934.2 $\pm$ 0.5 &    2 & 14 \\
1996.5 $\pm$ 0.6   & 1 & 15 \\
    2059.5 $\pm$ 0.4   & 0 & 16 \\
    2112.9 $\pm$ 0.3   & 1 & 16 \\
    2169.7 $\pm$0.3   & 2 & 16 \\
    2175.8 $\pm$0.5   & 0 & 17 \\
    2230.7 $\pm$0.3   & 1 & 17 \\
    2287.2 $\pm$0.3   & 2 & 17 \\
   2293.3 $\pm$0.3    & 0 & 18 \\
    2348.5$\pm$ 0.2   & 1 & 18\\
    2405.8$\pm$ 0.3   & 2 & 18\\
    2410.9$\pm$ 0.3   & 0 & 19\\
    2466.7$\pm$ 0.2   & 1 & 19\\
    2523.9$\pm$ 0.3   & 2 & 19\\
    2528.2$\pm$ 0.3   & 0 & 20\\
    2584.4$\pm$ 0.3   & 1 & 20\\
    2641.8$\pm$ 0.3   & 2 & 20\\
    2645.7$\pm$ 0.3   & 0 & 21\\
    2702.6$\pm$ 0.3   & 1 & 21\\
    2765.2$\pm$ 0.5   & 0 & 22\\
    2821.7$\pm$ 0.5   & 1 & 22\\
\tableline
\end{tabular}
\end{table}

We computed the models for Kepler-10 using two evolutionary codes. The first grid of models was calculated with the Aarhus Stellar Evolution Code \citep[ASTEC;][]{Christ2008a}. This uses the OPAL equation of state \citep[see][]{Rogers:1996iv} and OPAL opacities \citep{ir96}, combined with opacities from \citet{Ferguson:2005gn} at low temperature. ASTEC calculates nuclear reactions with the NACRE \citep{Angulo1999} parameters, the diffusion and settling of helium and heavy elements (using fully-ionized oxygen rates) with the approximations of \citet{Michau1993}, and convection with the \citet{Bohm1958} mixing-length formulation, characterized by a mixing length $\alpha_{\rm ML}$ in units of the local pressure scale height.

The resulting grid of models spans the range $0.8-0.98 \, M_\odot$ in steps of $0.01 \, M_\odot$, with $\alpha_{\rm ML} = 1.5, 1.8$ and $2.1$, and the nine combinations of composition corresponding to an initial hydrogen abundance by mass of $X_ 0 = 0.7246, 0.7299$ and $0.7346$ and initial heavy-element abundance of $Z_0 = 0.0111, 0.0127$ and $0.0144$. We used the Aarhus stellar pulsation code \citep[ADIPLS;][]{Christ2008b} to compute the frequencies of relevant models in the grid, corrected for the near-surface deficiencies in the modeling essentially following the procedure of \citet{Kjelds2008} but adapted to include modes of all degrees \citep[see][]{Silva2013}. We fit the observed frequencies similar to the method described in \citet{Gillil2013}, but including also [Fe/H] in the $\chi^2$ calculation. The inferred values for the stellar parameters obtained from this model can be seen in Table~\ref{table:modelpar} and the oscillation frequencies for the best-fitting model are marked with red open 
circles in the bottom panel of Figure~\ref{fig:power}.

The second grid of models was constructed using the Garching Stellar Evolution Code \citep[GARSTEC;][]{ws08}. The input physics includes OPAL \citep{ir96} and \citet{Ferguson:2005gn} opacities for high and low temperatures respectively, the \citet{gs98} solar mixture, and the 2005 version of the OPAL EOS \citep{Rogers:1996iv} complemented by the MHD EOS for low temperatures \citep{Hummer:1988kn}. Nuclear reaction rates are those from NACRE \citep{Angulo1999} with the updated cross section for $^{14}\mathrm{N}(p,\gamma)^{15}\mathrm{O}$ from \citet{Formicola:2004dl}. Convection is treated within the mixing-length formalism as described in \citet{kww13}, using the solar calibrated value for $\alpha_{\mathrm{ML}}=1.791$. Diffusion of helium and heavy elements is included using the \citet{at94} prescription.

The grid comprises masses between $0.85-0.95 M_\odot$, with the initial composition selected by exploring a range of initial helium abundances between $Y_\mathrm{i}=0.248-0.300$. For each of these initial values of helium, models for six metallicities that cover the 1\,$\sigma$ spectroscopic range of [Fe/H] are calculated. We computed frequencies for the models using ADIPLS, and calculate the $\chi^2$ fit to frequency combinations and spectroscopic data as described in \citet{Silva2013}. The final values are determined by a weighted average and standard deviation using the combined $\chi^2$ values. The oscillation frequencies obtained from the best-fitting GARSTEC model are marked by blue triangles in the bottom of Figure~\ref{fig:power}.

The inferred values for the stellar parameters obtained from these models are also given in Table~\ref{table:modelpar}. 
Although both determinations agree very well within their quoted errors, the method applied to the GARSTEC grid in determining stellar parameters is less sensitive to uncertainties arising from the poorly modeled stellar outer layers (see \cite{Silva2013} for an extended discussion). Thus, the stellar parameters used henceforward are determined by taking the GARSTEC values as the reference and adding in quadrature to the formal uncertainties the difference with respect to the ASTEC results, as found in Table~\ref{table:modelpar}.
\begin{table}[ht]
\centering
\footnotesize
\caption{Stellar parameters of the host star}
\label{table:modelpar}
\begin{tabular}{l c c}
\tableline\tableline
Parameter				& ASTEC				& GARSTEC   \\
\tableline
Radius, $R_*~(R_\sun)$	& 1.062 $\pm$ 0.009	& 1.065 $\pm$ 0.008	\\
Mass, $M_* ~(M_\sun)$	& 0.912 $\pm$ 0.023	& 0.913 $\pm$ 0.022	\\
Density, $\bar{\rho}_*$ (g/cm$^3$)	& 1.071 $\pm$ 0.003 & 1.068 $\pm$ 0.002	\\
Age (Gyr)				& 10.50 $\pm$ 1.33	& 10.41 $\pm$ 1.36	\\
Luminosity				& 1.032 $\pm$ 0.063	& 1.091 $\pm$ 0.079	\\
\tableline
\end{tabular}
\end{table}
\section{Planetary Transit Analysis}\label{sec:transit}
As for the asteroseismic analysis (Section~\ref{sec:asteroseismology}), we exclusively use the available short-cadence observations for the planetary analysis. We again start out from raw data (simple aperture photometry), to avoid that prior data corrections affect the analysis of the planetary phase curve (see Section~\ref{sec:phasecurve} for details).

For both Kepler-10b and Kepler-10c, we first determine the time of central transit for each individual transit throughout the whole time series. This is done by fitting a transit curve to each individual transit. The orbital period is determined by fitting a straight line to the transit times vs. the transit numbers, assuming a constant orbital period. The assumption of constant period is verified by investigating possible transit timing variations (TTVs), using a power spectrum of the O-C (observed - calculated) residuals, which could indicate a periodic signal from a perturbing object. We found no evidence of TTVs and place an upper limit of 2 minutes on the amplitude of any sinusoidal variations. The periods are listed in Table \ref{table:parameters}.

We analyzed the phase-folded and binned data using the Transit Analysis Package (TAP) described in \cite{2012AdAst2012E..30G}. TAP uses Markov Chain Monte Carlo (MCMC) simulations to fit the transit model with the analytical methods described by \cite{2002ApJ...580L.171M}. The model outputs from the MCMC analysis are inclination, $i$, scaled semi-major axis, $a/R_\star$, relative radius, $r_\mathrm{p}/R_\star$, as well as linear and quadratic limb-darkening (LD) coefficients. We leave the LD coefficients as free fitting parameters, but they are unique to both transit fits as they are fitted simultaneously.

We assessed the circularity of orbits of Kepler-10b and -10c by comparing the stellar densities calculated from transit parameters to those determined by other means as described in \cite{tingley2011}. In this case, we used the value for stellar density obtained by asteroseismic analysis. However, this means that we cannot use the transit parameters listed in Table~\ref{table:parameters}, as they used the asteroseismic stellar density as a prior and fixed the orbit to be circular. This would mean that the fitting algorithm would be naturally biased towards solutions that return circular orbits. Therefore, in order to assess an unknown eccentricity, it is important to use a fitting approach that only fits the transit and does not attempt to fit orbital parameters as well. We therefore used transit parameters given in Rowe et al. (submitted), which are fit in this manner.  We find that the minimum eccentricities for both planets are consistent with 0: $0.050^{+0.012}_{-0.050}$ for Kepler-10b and  $0.042^{+0.015}_{-0.042}$ for Kepler-10c, which supports the use of the asteroseismic stellar density as a prior in the transit fits (a correction term which depends on the eccentricity and angle of periastron would have to be included in case of an eccentric orbit). We used this eccentricity information along with the derived period and the stellar mass to recalculate the mass of Kepler-10b using the radial-velocity semi-amplitude that was observed by \citet{2011ApJ...729...27B}; see Table~\ref{table:parameters}.
 
Running the TAP routine provides poor constraints on the inclination angle for both exoplanets. For instance, in the Kepler-10b case the output values are $i = 83.6 ^{+3.8}_{-2.4}(^\circ)$, $a/R_\star = 3.38 ^{+0.21}_{-0.18}$, and $r_\mathrm{p}/R_\star = 0.01260 ^{+0.00021}_{-0.00022}$. However, the above results on the orbital eccentricity suggest that we could use the asteroseismic density as a prior when fitting the planetary transit: the mean stellar density ($\bar{\rho}_\star$) yields $a/R_\star$ through Kepler's third law (and assuming $M_\star \gg m_\mathrm{p}$):
\begin{equation}
\bar{\rho}_\star 
= \frac{3\pi}{GP^2}
\left(\frac{a}{R_\star}\right)^3,
\label{eq:astero_density}
\end{equation}
where $P$ is the orbital period, and $G$ the gravitational constant. The asteroseismic analysis of the stellar p-modes yields a stellar mean density of $1.068\pm0.004$ g/cm$^3$, which results in $\left(a/R_\star\right)_\mathrm{astero} = 3.408\pm0.004$ and $\left(a/R_\star\right)_\mathrm{astero} = 48.75\pm0.06$, for Kepler-10b and Kepler-10c, respectively. As can be seen, the inclusion of a fixed $\left(a/R_\star\right)_\mathrm{astero}$ significantly improves the error bars on the orbital inclination. The final values are listed in Table~\ref{table:parameters}.
\section{Analysis of Phase Curve \& Occultation Depth}\label{sec:phasecurve}

Because of the short period of Kepler-10b, we can look at flux variations throughout the entire phase curve for this planet. We calculate this curve in a two-step procedure. We first de-trend the whole time series by using a moving-median-filtered time series where the length of the filter is set to the planetary period for Kepler-10b. We also calculated de-trended time-series data using filter lengths of two and three times the planetary period for comparison, which revealed no clear dependence on the length of the filter. The second part of the analysis includes first a phase folding of the time series using the planetary period. 

We calculate the median filter in smaller phase bins, using the median value in order to remove extreme data points which could affect the mean values. For Kepler-10b, each bin includes about 2000 individual data points.
The final data set we use consists of 603 data points, which means that the mean bin width is 2 minutes, which is two times the temporal SC resolution.

\begin{figure*}[ht]
\centering
\includegraphics[width = \textwidth]{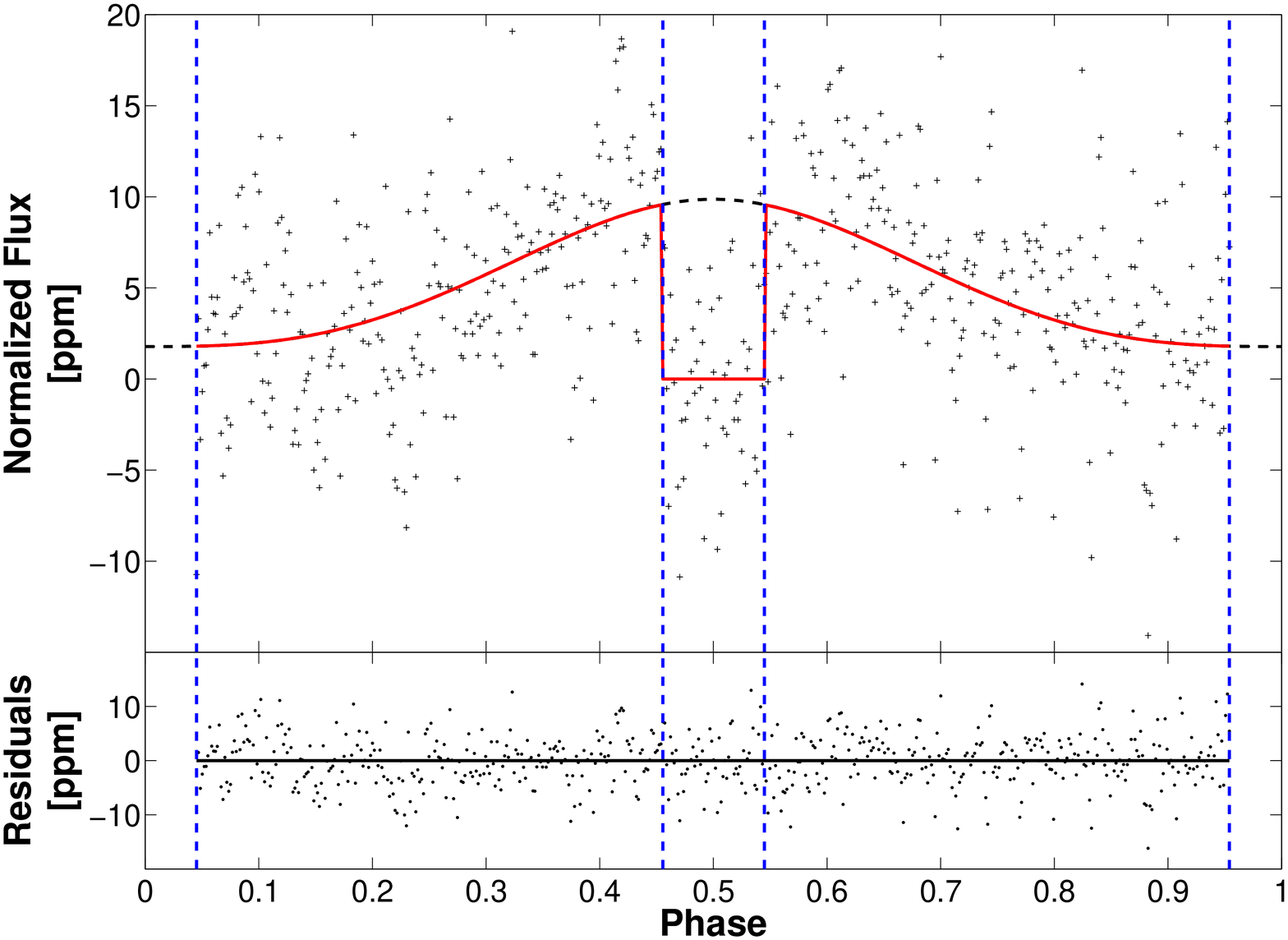}
\caption{\textit{Top:} Phase-folded and binned observations of the Kepler-10 system (black crosses). The reflective/thermal part of Eq.~\ref{eq:fit_eq} is shown as the dashed black line. The red curve shows the best model fit for Eq.~\ref{eq:fit_eq}, setting the ellipsoidal and beaming terms $A_\mathrm{e} = A_\mathrm{b} = 0$. Inside the occultation, a simple flat bottom is fitted to the data. \textit{Bottom:} Residuals. The vertical dashed blue lines separate the data that are fitted using Eq.~\ref{eq:fit_eq} from the occultation data.}
\label{Phase_fit}
\end{figure*}

The phase-folded data of the Kepler-10b system shows small-scale variations $\delta$ in flux, throughout the entire phase $\phi$ of the planet, as can be seen in Figure~\ref{Phase_fit}. These variations are due to a combination of reflection plus thermal component from the planet, ellipsoidal distortions of the star, and Doppler beaming \citep{2012ApJ...761...53B}. Assuming that the planet behaves as a Lambertian sphere \citep{russell1916}, these variations can be described by
\begin{eqnarray}
\delta = \frac{\Delta F}{F} = C - A_\mathrm{e}\mathrm{cos}(4\pi\phi) + A_\mathrm{b}\mathrm{sin}(2\pi\phi) \nonumber\\
- A_\mathrm{r}\frac{\mathrm{sin}(z) + (\pi - z)\mathrm{cos}(z)}{\pi}\ ,
\label{eq:fit_eq}
\end{eqnarray}
where $A_\mathrm{e}$, $A_\mathrm{b}$ and $A_\mathrm{r}$ are the ellipsoidal, beaming and reflective plus thermal emission amplitudes, respectively, and $C$ is an arbitrary vertical offset. The parameter $z$ is defined as
\begin{equation}
\cos(z) = -\sin(i)\cos(2\pi \phi).
\end{equation}
The expected amplitudes for the beaming and ellipsoidal variations
are very small. These amplitudes can be estimated using the analytical models shown in
\cite{2012ApJ...751..112J}:
\begin{eqnarray}
A_\mathrm{e} = q \cdot \sin^2(i) \cdot a^{-3},\\
A_\mathrm{b} = 4 \cdot \frac{K_z}{c}.
\end{eqnarray}
Using the values for the mass-ratio, $q$, and semi-velocity amplitude, $K_z$, 
from \cite{2011ApJ...729...27B} and the values found in Table~\ref{table:parameters} yields $A_\mathrm{e} \approx 5 \cdot 10^{-4}$ parts per million (ppm) and $A_\mathrm{b} \approx 0.04$~ppm. The theoretical amplitudes are small and with a mean data scatter of $\sim4.8$~ppm, we do not expect them to be measurable. After confirming that our fitting routine could indeed not constrain those amplitudes, we therefore do not include the beaming and ellipsoidal terms into the final phase model.
Therefore, our final fit includes only the reflective plus thermal emission (Lambertian) component of Eq.~\ref{eq:fit_eq}, yielding $A_\mathrm{r} = 8.13\pm0.68$ ppm. All data points outside the transit and occultation are used to determine this amplitude, where the occultation is taken to be centered at mid-phase and of the same duration as the transit.

The flux level of data points inside the occultation can then be measured as well. When measuring the occultation depth it is necessary to be specific on how exactly this is defined. From the observed phase curve, one may calculate the difference between the flux level at the wings of the phase curve (just before and after the occultation), and the flux level during the occultation. However, if we use this as the measured occultation depth, the measurement depends on the inclination of the system and more importantly on the distance between the star and the planet (which determines when the occultation starts and ends, e.g. if the planet is close to the star the occultation will start long before the planet reaches the point of superior conjunction, where the disk is fully illuminated). If the inclination is different from 90 degrees, we will find that superior conjunction does not correspond to a fully illuminated disk. Similarly, at inferior conjunction, we will observe a planetary disk that has a tiny 
illumination at one limb and we are therefore not only measuring the light emitted from the nightside.

\begin{figure*}[ht]
\centering
\includegraphics[width = \textwidth]{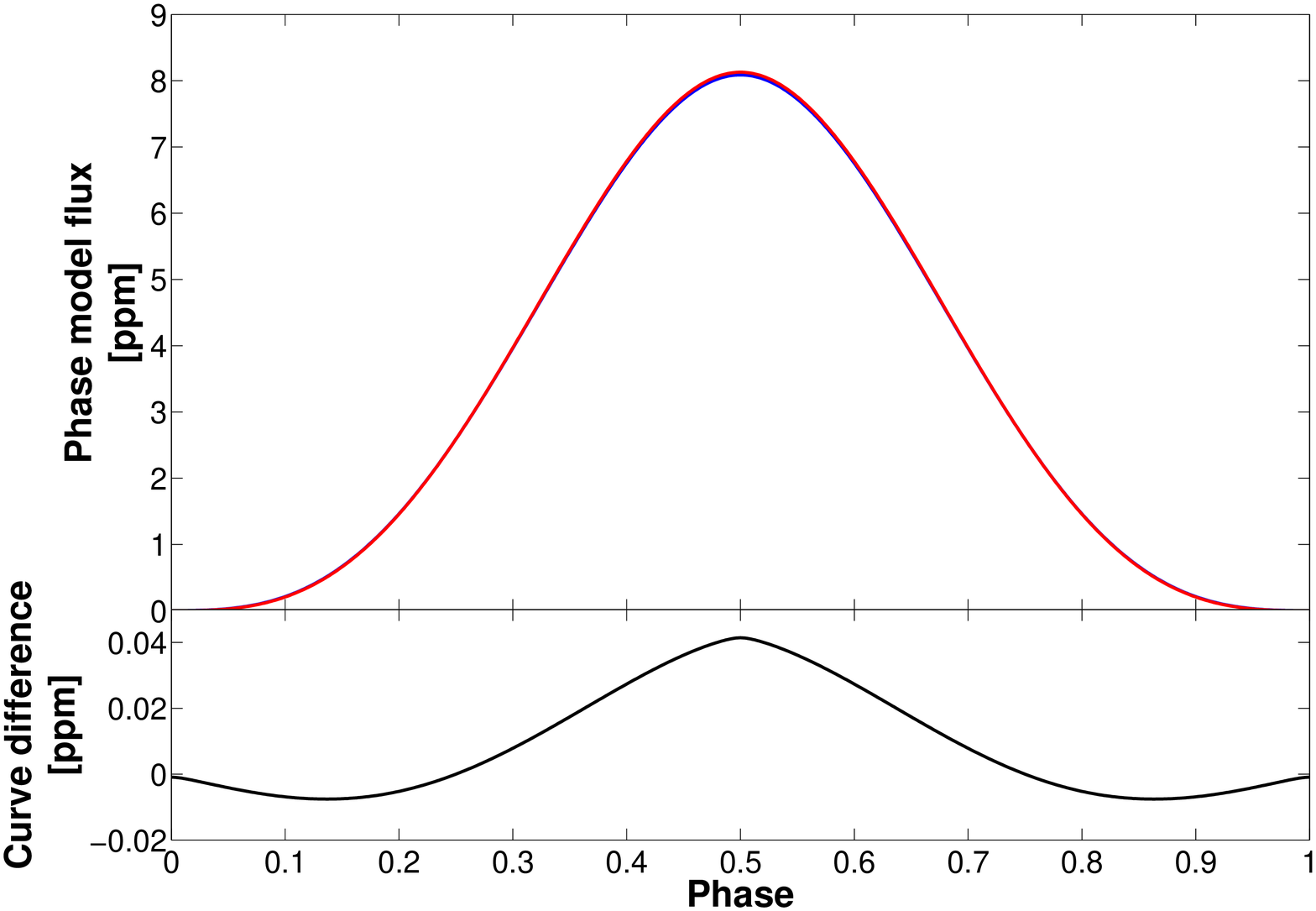}
\caption{\textit{Top:} Shown are the two phase models described by Eq.~\ref{eq:fit_eq} using the listed $A_\mathrm{r}$-value, where the blue curve is using $i = 84.15^\circ$ and the red curve uses $i = 90^\circ$. 
\textit{Bottom:} The difference between the two phase-models. The largest deviation is seen at superior conjunction (phase = 0.5).}
\label{Phase_difference}
\end{figure*}

In order to provide a physical meaning to the occultation depth, we will use the model presented in Eq.~\ref{eq:fit_eq} to determine the flux level at superior conjunction, for a planet at an inclination of $90^\circ$, and compare this with the measured flux level inside the occultation. Similarly, the nightside flux contribution can be measured by comparing the occultation flux level with the model at inferior conjunction and corrected for the inclination angle and 
position of superior conjunction (0.04 and 0.27 ppm, 
respectively). The latter two effects correct for the fact 
that the exoplanet is not fully illuminated. The effect of 
using $i = 84.15^\circ$ and $i = 90^\circ$ can be seen on 
Figure~\ref{Phase_difference}, where the difference between the two phase models is shown. At superior conjunction the change corresponds to 0.04 ppm, while the change at inferior conjunction is approximately $10^{-3}$ ppm. This is well within our error 
bars; we include it only for the sake of completeness. 
With these definitions, we measure $\delta_\mathrm{occultation} = 9.91\pm1.01$ ppm and $\delta_\mathrm{night} = 1.78\pm0.76$ ppm.

The fit to the phase curve and the occultation, as well as the residuals, are shown in Figure~\ref{Phase_fit}.
\section{Planetary temperature and albedo}\label{sec:albedo}
The planetary phase curve and occultation can be used to study Kepler-10b's temperature and albedo. We approximate both star and planet as black bodies and use Planck's law to describe the photon flux (amount of photons measured) per unit wavelength:
\begin{equation}
 \tilde{B}_\lambda(T) = \frac{2c}{\lambda^4} \frac{1}{\exp\left(\frac{hc}{\lambda k_b T}\right) - 1} \label{eq:planck_law},
\end{equation}
where $c$ is the speed of light, and $h$ and $k_b$ are the Planck and Boltzmann constants. The amount of photons emitted at a wavelength $\lambda$ is then only dependent on the temperature $T$ of the object. We approximate the stellar temperature ($T_\star$) as the spectroscopically determined effective temperature \citep[][see Table~\ref{table:parameters}]{torres2012}. We integrate the Planck function over the \textit{Kepler} bandpass, using the \textit{Kepler} instrument response function that is available for download\footnote{http://keplergo.arc.nasa.gov/kepler\_response\_hires1.txt} from the \textit{Kepler} website.

Even if a planet might not behave as a black body, its brightness temperature ($T_\mathrm{p,b}$) in the \textit{Kepler} bandpass can be defined as the temperature of a black body with equivalent flux in the \textit{Kepler} band. This temperature can be estimated from the differences in flux $\delta_\mathrm{flux}$ that may be present at different parts of the phase curve, assuming that they arise due to the thermal radiation of the planet:
\begin{equation}
 \delta_\mathrm{flux} = \left(\frac{R_\mathrm{p}}{R_\star}\right)^2 \frac{\tilde{B}_\lambda(T_{\mathrm{p,b}})}{\tilde{B}_\lambda(T_\star)}. \label{eq:thermal_light}
\end{equation}
This equation can be directly used to measure the night-side temperature of the planet, by comparing the flux at the wings of the transit with the flux level inside the occultation. Setting $\delta_\mathrm{flux} = \delta_\mathrm{night}$, and using Table~\ref{table:parameters}, we find $T_{\mathrm{p,b}} \mathrm{(night)} = 2600_{-175}^{+124}$ K. We caution that the error bars can be misleading as they are highly asymmetric: the measurement of $\delta_{\mathrm{night}}$ is zero within $2.3\sigma$, and that would imply a zero nightside temperature. However, once a non-zero nightside flux is measured, a high brightness temperature is immediately required to explain this.

Following the same strategy, the temperature of the planetary dayside can be calculated. However, here the situation is more complicated: the photon flux from the dayside can be a consequence of thermal radiation, but could also be caused by reflection of stellar flux. The reflection is measured by the planetary geometric albedo, $A_\mathrm{g}$:
\begin{equation}
 \delta_\mathrm{flux} = \left(\frac{R_\mathrm{p}}{R_\star}\right)^2 \frac{\tilde{B}_\lambda(T_{\mathrm{p,b}})}{\tilde{B}_\lambda(T_\star)} + A_\mathrm{g} \left(\frac{R_\mathrm{p}}{a}\right)^2. 
 \label{eq:reflected_light}
\end{equation}

\begin{figure*}[ht]
\centering
\includegraphics[width = 0.8\textwidth]{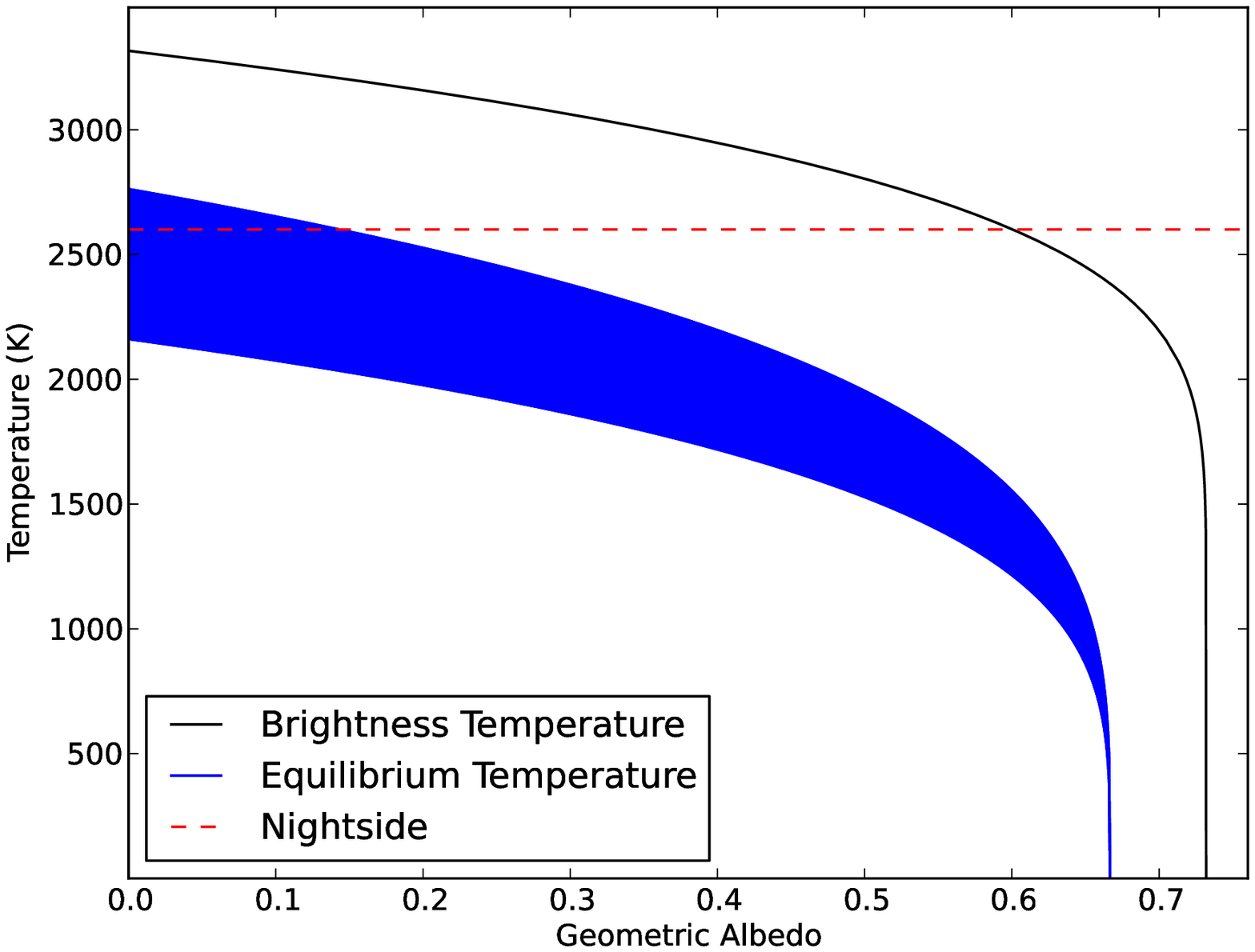}
\caption{The black line represents the brightness temperature of the planetary dayside $T_\mathrm{p,b} (\mathrm{day})$ as a function of the geometric albedo $A_\mathrm{g}$. The maximum temperature is found for zero albedo (pure emission), a maximum albedo corresponds to temperature zero (pure reflection). The blue area shows the equilibrium temperature for different levels of reflection, and for all scenarios between uniform heat redistribution (lowest equilibrium temperature) and no heat redistribution (highest equilibrium temperature). The equilibrium temperature is lower than the brightness temperature in all cases. This could indicate internal heating from the planet, or excess planetary flux relative to a blackbody in the \textit{Kepler} bandpass. The dotted red line shows the measured night-side temperature ($T_{\mathrm{p,b}}$(night)).}
\label{fig:temperature_albedo}
\end{figure*}

Eq.~\ref{eq:reflected_light} can then be used to understand the planetary dayside. The flux received from the planetary dayside (the occultation depth $\delta_\mathrm{occultation}$) can be attributed to a combination of thermal emission (a function of temperature) and reflected stellar light (a function of albedo). Assuming all day-side flux comes from emission ($A_\mathrm{g} = 0$), a maximum day-side brightness temperature is estimated as $T_{\mathrm{p,b}} (\mathrm{day}) = 3316_{-56}^{+52}$ K. The other extreme scenario corresponds to pure reflection ($T_\mathrm{p,b} (\mathrm{day}) = 0$); a maximum albedo $A_\mathrm{g} = 0.73$ is found. A curve relating temperature and albedo between these two extreme scenarios is plotted in Figure~\ref{fig:temperature_albedo}.

The observed temperature-albedo measurements can be compared to a theoretical temperature. The equilibrium temperature ($T_\mathrm{eq}$) is the temperature of an isothermal planet with no internal heating. It depends on the amount of radiation it receives from the star, the amount of light reflected and an internal heat redistribution parameter $f$ \citep[e.g.][]{lopezmorales2007}:
\begin{equation}
 T_\mathrm{eq} = T_\star \left(\frac{a}{R_\star}\right)^{-1/2} \left(f\left(1-\frac{3}{2}A_\mathrm{g}\right)\right)^{1/4},
\end{equation}
assuming that the geometric albedo, $A_\mathrm{g}$, is 2/3 of the planetary Bond albedo. The heat redistribution parameter as defined here can range from 1/4 to 2/3, with the former value representing homogeneous redistribution of incident radiation around the planet such that the night-side temperature equals the day-side  and the latter number representing no redistribution, or instant re-radiation. The true value of $f$ presumably lies somewhere between those two scenarios. As before, we can calculate maximum temperatures for those two scenarios, assuming $A_\mathrm{g} = 0$, and find $T_\mathrm{eq,rerad} = 2760$ K and $T_\mathrm{eq,hom} = 2160$ K. The full region of equilibrium temperatures for different redistribution factors and different albedos is shown in Figure \ref{fig:temperature_albedo}.

We note that the equilibrium temperature is lower than the brightness temperature in all cases. One explanation for this would be an additional source heating the planet, which we do not take into account (e.g. internal heating from the planet). Another explanation is that the excess flux in the \textit{Kepler} bandpass is caused by a deviation of the planetary spectrum from a black-body spectrum. Interestingly, \citet{esteves2013} analyze phase curves and occultation for eight Hot Jupiter planets in the \textit{Kepler} field and find a similar result.
\section{Discussion and conclusion}
We revisited the analysis of the Kepler-10 system, as the amount of data available has significantly increased since the discovery paper. We derive system parameters (Table \ref{table:parameters}) to an unprecedented accuracy. At this level, a word of caution about systematic effects is in order. For the host star, the two independent asteroseismic analyses agree within the error bars for all parameters, supporting the accuracy of our final values. At the planetary level, the accuracy of order 1\% for the relative planetary radius is at the level of expected \textit{Kepler} systematic effects \citep{vaneylen2013}. Finally, the planetary phase curve is measured at the part per million level and therefore particularly vulnerable to data systematics or the choice of model. Particular care is needed when interpreting the temperature of the planet's nightside, where even a change at the ppm level can produce a very different temperature and be the difference between detection and non-detection of nightside radiation.

It is interesting that the parameters where our analysis provides the largest improvement (stellar age, planetary nightside temperature, and planetary dayside temperature plus albedo) are also those that are potentially the most relevant for models of the evolution and composition of Kepler-10b. We can now conclusively claim that the age of the system is on the order of 10 Gyr, with the value determined to a precision better than 15\%. The  parameters of the host star are consistent with the earlier analysis of \citet{2011ApJ...729...27B}, telling us that it is possible to get accurate stellar parameters with only a few months of data. However, our analysis drastically reduced the level of uncertainties.

Together with the recently discovered Kepler-78 \citep{sanchis2013}, Kepler-10b is the only rocky planet for which a phase curve has been detected. We analyzed it in much more detail than in the discovery paper \citep{2011ApJ...729...27B} and a subsequent physical model for the surface \citep{rouan2011}, which used the same limited dataset. 

Our values for $A_\mathrm{r}$ and $\delta_\mathrm{occultation}$ ($8.13\pm0.68$ and $9.91\pm1.01$) are a significant refinement of the values found by \cite{2011ApJ...729...27B}, $7.6\pm2.0$ and $5.8\pm2.5$ ppm respectively, which were based on about five months of \textit{Kepler} data. \cite{rouan2011}, using the same dataset, find a single-parameter model in which phase curve amplitude and occultation depth are equal, and they find a value of $5.6\pm2.0$ ppm.
While consistent with these values within $2\sigma$, our values are on the high end for both, and determined more accurately. In addition, our non-zero value for $\delta_\mathrm{night}$ is different from what was used in the Lava-Ocean model by \cite{rouan2011}, who assess a model where the planet's surface is filled with molten lava, which is not expected to flow over to the nightside (see \cite{leger2011}). 
Our apparent non-zero value leads to a high temperature for the planetary nightside.
We note that the measurement of this temperature might depend on the model used for the phase curve. 
To check the contribution of the $\delta_\mathrm{occ}$ and $\delta_\mathrm{night}$ in the phase model, we estimate both $\delta_\mathrm{occ}$ and $\delta_\mathrm{night}$, by
comparing the continuum data surrounding the occultation 
and primary transit event to the occultation level. 
For consistency, the continuum bins consist of the same number
of data points as is contained in the occultation event.
This basic method yields values of $\delta_\mathrm{occ} = 8.74\pm1.47$ ppm and
$\delta_\mathrm{night} = 2.80\pm1.51$ ppm.
The respective decrease and increase of $\delta_\mathrm{occ}$ and $\delta_\mathrm{night}$
are not surprising as the basic methodology does not take into account
the curvature of the phase curve.
Future work might be warranted to improve on this and confirm the nightside temperature.
\begin{table*}[th]
\centering
\caption{Measured parameters of Kepler-10b.
References:~a.~\citet{torres2012}, b.~\citet{2011ApJ...729...27B}} 
\label{table:parameters} 
\centering 
\begin{tabular}{l c}
\tableline\tableline 
Parameter & Value \\ 
\tableline 
\textit{Stellar parameters} & \\
Mass, $M_\star\ (M_\sun)$ & $0.913 \pm 0.022$\\
Radius, $R_\star\ (R_\sun)$ & $1.065 \pm 0.009$\\
$\bar{\rho}_\star$ (g/cm$^3$) & $1.068 \pm 0.004$\\
Age (Gyr) & $10.41 \pm 1.36$\\
Temperature, $T_\star$ (K) & $5643 \pm 75^\mathrm{a}$\\
Spectroscopic metallicity, [Fe/H]	& $-0.15 \pm 0.07^\mathrm{a}$\\
\hline
\textit{Transit and orbital parameters: Kepler-10b} & \\
Orbital period, $P$ (days)  &	$0.8374912 \pm 0.0000003$ \\
Scaled semimajor axis, $a/R_\star$ & $3.408 \pm 0.004$\\
Scaled planet radius, $R_\mathrm{p}/R_\star$ & $0.01254 ^{+0.00013}_{-0.00013}$\\
Orbital inclination, $i$ ($^\circ$) & $84.15 ^{+0.17}_{-0.18}$\\
Linear LD & $0.21^{+0.11}_{-0.10}$	 \\
Quadratic LD & $0.63^{+0.15}_{-0.16}$ \\
Eccentricity & $0$ \\
Orbital semi-amplitude $K$ (m s$^{-1}$) & $3.3\pm0.9^\mathrm{b}$ \\
\hline
\textit{Derived planetary parameters: Kepler-10b} & \\
Orbital semi-major axis, $a$ (AU) & $0.01687\pm0.00014$ \\
Radius, $R_\mathrm{p}$ ($R_\oplus$) & $1.46\pm0.02$ \\
Mass, $M_\mathrm{p}$ ($M_\oplus$) & $4.60\pm1.26$ \\
Density, $\rho_\mathrm{p}$ (g cm$^{-3}$) & $8.15\pm2.26$ \\
\hline
\textit{Observed phase curve parameters: Kepler-10b} & \\
$\delta_\mathrm{occultation}$ (ppm) & $9.91\pm1.01$\\
$A_\mathrm{r}$ (ppm) & $8.13\pm0.68$\\
$\delta_\mathrm{night}$ (ppm) & $1.78\pm0.76$\\
\hline
\textit{Derived phase curve parameters: Kepler-10b} &\\
$A_\mathrm{g}$ (max) & $0.73$ \\
$T_{\mathrm{p,b}}$ (day,max) (K)& $3316_{-56}^{+52}$\\
$T_{\mathrm{p,b}}$ (night) (K)& $2600_{-175}^{+124}$\\
$T_\mathrm{eq,rerad}$ (K) & $2760$\\
$T_\mathrm{eq,hom}$ (K) & $2160$\\
\hline
\textit{Transit and orbital parameters: Kepler-10c} & \\
Orbital period, $P$ (days)  &	$45.29427\pm0.00012$ \\
Scaled semimajor axis, $a/R_\star$ & $48.75\pm0.06$\\
Scaled planet radius, $R_\mathrm{p}/R_\star$ & $0.01998^{+0.00020}_{-0.00021}$\\
Orbital inclination, $i$ ($^\circ$) & $89.640^{+0.023}_{-0.017}$\\
Linear LD & $0.21^{+0.11}_{-0.10}$	 \\
Quadratic LD & $0.63^{+0.15}_{-0.16}$\\
$e_{\rm min}$ & $0.025^{+0.027}_{-0.025}$\\
\hline
\textit{Derived planetary parameters: Kepler-10c} & \\
Orbital semi-major axis, $a$ (AU) & $0.2413\pm0.0021$ \\
Radius, $R_\mathrm{p}$ ($R_\oplus$) & $2.32\pm0.03$ \\
\tableline 
\end{tabular}
\end{table*}
\acknowledgments
Funding for this Discovery mission is provided by NASA's Science Mission Directorate. The authors wish to thank the entire Kepler team, without whom these results would not be possible. Funding for the Stellar Astrophysics Centre is provided by The Danish National Research Foundation (grant agreement No. DNRF106). The research is supported by the ASTERISK project (ASTERoseismic Investigations with SONG and Kepler) funded by the European Research Council (grant agreement No. 267864).
The authors wish to thank the referee, Ron Gilliland, for his suggestions, which have substantially improved the paper. 

\end{document}